# Recent Results from the ANTARES Neutrino Telescope

**Giorgio Giacomelli, for the Antares Collaboration**
University of Bologna and INFN Sezione di Bologna
Viale Berti Pichat 6/2, 40127 Bologna, Italy, e-mail: giacomelli@bo.infn.it

**Abstract.** The ANTARES underwater neutrino telescope is located in the Mediterranean Sea about 40 km from Toulon at a depth of 2475 m. In its 12 line configuration it has almost 900 photomultipliers in 295 "floors". The performance of the detector is discussed and several results are presented, including the measurements of downgoing muons, atmospheric neutrinos, search for a diffuse flux of high energy muon neutrinos, search for cosmic point sources of neutrinos, multi messenger astronomy, searches for fast magnetic monopoles and slow nuclearites. A short discussion is also made on Earth and Sea Science studies with a neutrino telescope.

*Keywords*: Neutrino telescope, cosmic point sources, diffuse neutrino flux, dark matter, monopoles, nuclearites

1. INTRODUCTION

The effort to build large sea water Cherenkov detectors was pioneered by the Dumand Collaboration with a prototype at great depths close to the Hawaii islands [1]; the project was eventually cancelled. Then followed the fresh water lake Baikal detector at relatively shallow depths [2]. Considerable progress was made by the AMANDA and IceCube telescopes in Antarctica [3,4]. In the Mediterranean Sea the NESTOR collaboration tested a deep line close to the Greek coast [5] and the NEMO Collaboration tested a number of prototypes close to Sicily [6]. These groups plus ANTARES [7] are now involved in a major project, Km3Net, aimed to the construction of a 1 km$^3$ size detector in the Mediterranean Sea [8].

ANTARES is a deep sea neutrino telescope, mainly designed for the detection of high-energy neutrinos emitted by astrophysical sources, galactic and extragalactic [9]. The telescope is also sensitive to neutrinos produced via dark matter annihilation inside massive bodies like the Sun and the Earth. Other physics topics include atmospheric neutrinos, searches for fast magnetic monopoles and slow nuclearites [10]. ANTARES studies neutrinos from the southern hemisphere, which includes neutrinos from the center of our Galaxy. It is thus complementary to the studies by the ice detectors at the South Pole [4]. ANTARES is also a unique deep-sea marine observatory, providing continuous monitoring with a variety of sensors dedicated to Oceanographic and Earth Science studies [11].

2. THE ANTARES DETECTOR

The ANTARES neutrino telescope is located in the Mediterranean Sea, about 40 km from the coast of Toulon, France, at 42°48'N, 6°10'E, and at a depth of 2475 m [7] [12,13,14]. A schematic view of the detector is shown in Fig. 1 [7,9]. The detector is an array of photomultiplier tubes (PMTs) arranged on 12 flexible lines. Each line has up to 25 detection storeys of triplets of optical modules (OMs), each equipped with three downward looking 10 inch photomultipliers oriented at 45° to the line axis [15,16]. The lines are kept vertical by a buoy at the top of the 450 m long line. The spacing between storeys is 14.5 m and the lines are spaced by 60-70 m and are on an octagonal structure [7].

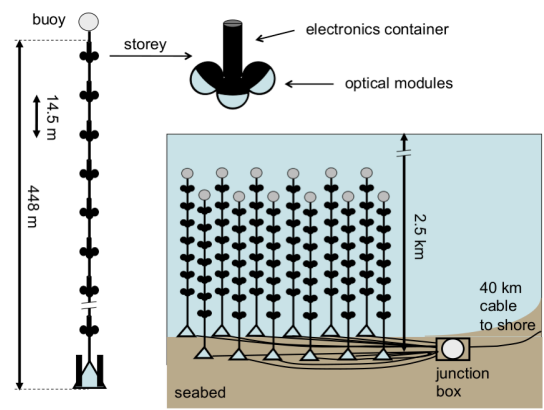

**Fig. 1**. The layout of the 12 line ANTARES detector. The sketches at the left and at the top show some details.

An acoustic positioning system provides real time location of the detector elements with a precision of few centimeters [17,18]; a system of optical beacons allows in situ calibrations [19]. Since may 2008 the detector is running in its final configuration of 12 lines. An additional line (IL07) contains oceanographic sensors dedicated to the measurement of environmental parameters [20,21]. Line 12 and IL07 have hydrophone storeys to make calibrations and to measure the ambient acoustic backgrounds.

The lines are connected to the Junction Box (JB) that distributes power and data from/to shore. The instrumented part of each line starts at 100 m above the sea floor, so that Cherenkov light can be seen also from upgoing muons coming from neutrino interactions in the rock below the sea or in the sea water beneath the Photomultipliers (PMTs). The three dimensional structure of PMTs allows to measure the arrival time and position of Cherenkov photons produced by relativistic charged particles in the sea water. A reconstruction algorithm gives the direction of a muon, infer that of the incident neutrino and allows to distinguish upgoing muons, produced by neutrinos, from the more abundant downgoing muons, produced by cosmic ray interactions in the atmosphere.

The data acquisition is based on the "all data to shore" concept: all hits above a threshold of 0.3 single photoelectrons are digitized and sent to shore, where a computer farm applies a trigger requiring the presence of few coincidences between pairs of PMTs within a storey [22,23]. This typical trigger rate is 5-10 Hz, dominated by downgoing muons.

For neutrino energies above few TeV the angular resolution for the search of astrophysical point sources is determined by the timing resolution and accuracy of the location of the photomultipliers. The relative time calibration is performed by many systems: the measurement of the transit time of the clock signals to the electronics in each storey and the determination of the residual time offsets within each storey. The positions of the PMTs are measured by an acoustic positioning system and an optical beacon [19]. Fig. 2 shows the Sky observable from the South Pole and from the Mediterranean sea.

The energy measurement relies on an accurate calibration of the charge detected by each PMT and is computed via high enery models.

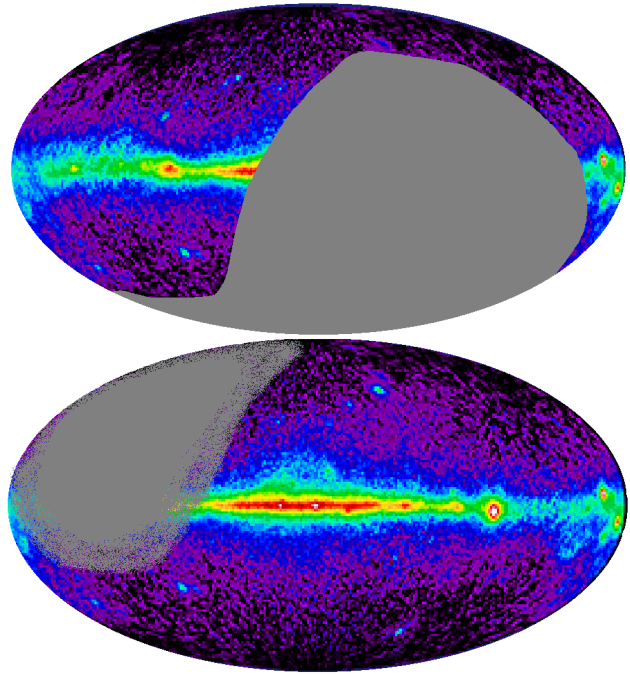

Fig. 2. Top: Sky observable by neutrino telescopes at the South Pole (Amanda, IceCube), Bottom: in the Mediterranean Sea (Antares) (Galactic coordinates).

3. ATMOSPHERIC MUONS

The main signal observed by ANTARES is due to downgoing atmospheric muons, whose flux exceeds that of neutrino induced muons by many orders of magnitude. They are produced by high energy cosmic rays interacting with atomic nuclei of the upper atmosphere producing charged pions and kaons, which then decay into muons and neutrinos.

Atmospheric muons are used to check the detector response and to test different Monte Carlo (MC) simulations. They are also important to measure the attenuation of the muon flux as a function of depth, to obtain the muon vertical depth intensity [24,25,26], and to verify the pointing accuracy of the detector (via the shadow of the cosmic ray flux by the moon [8]).

Fig. 3 shows one muon neutrino event reconstructed by the ANTARES on line reconstruction algorithm. Fig. 4 shows the ANTARES muon vertical intensity data [26] together with previous data [4,5] [27,28]. Fig. 5 shows a preliminary measurement of the Moon shadow. Many measurements by past experiments were used to establish their pointing accuracy [29].

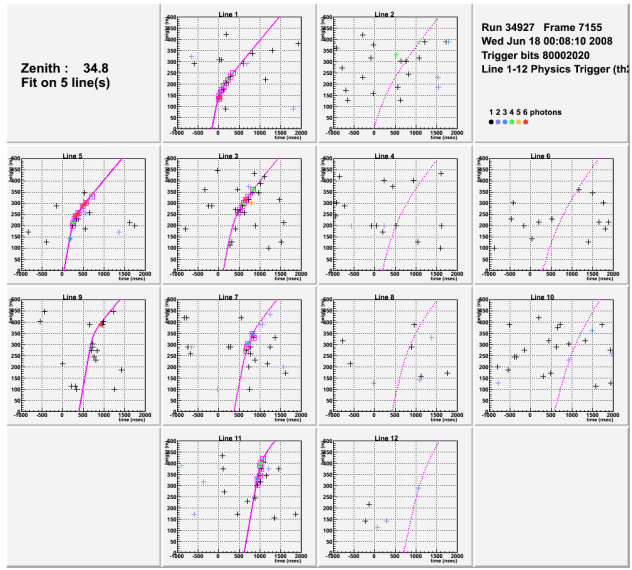

Fig. 3. Example of a neutrino candidate event seen in 5 ANTARES lines.

### 4. NEUTRINO ASTRONOMY. "POINT SOURCES"

The main reason to build neutrino telescopes is to study high energy muon neutrino astronomy. Neutrinos may be produced in far away sources, where charged and neutral pions are produced and decay. The neutrinos from charged pion decay reach the Earth, while the *photons* from $\pi^0$ decay may interact with the Cosmic Microwave Background radiation and with matter, *protons* are deflected by magnetic fields and *neutrons* are unstable. The main drawback of *neutrinos* is that one needs very large detectors [8,9]).

The muons produced in neutrino interactions can be distinguished from atmospheric muons requiring that they are upgoing. Fig. 3 shows one neutrino event which can be easily distinguished from a downgoing muon. Subsequent probability cuts allow to improve the separation. On the basis of the 2007+2008 data using a probability cut at $\Lambda > -5.4$ one had a neutrino sample with which was made a preliminary sky map [9,30].

A second study used the 2007-2010 data, making more stringent cuts : upgoing muons were required to have a zenith angle <90 degrees and quality parameter $\Lambda > -5.2$ [10.I](it means Ref[10] Sect. I). In this case the number of events is 3058 [10.II] (Ref 10 Sect. II ). Most of these events are from atmospheric muon neutrinos, which are an unavoidable background; the atmospheric muon contamination was smaller than 10%. Atmospheric muons were simulated with the MUPAGE package; neutrinos were generated with the GENNEU package and the Bartol model [26].

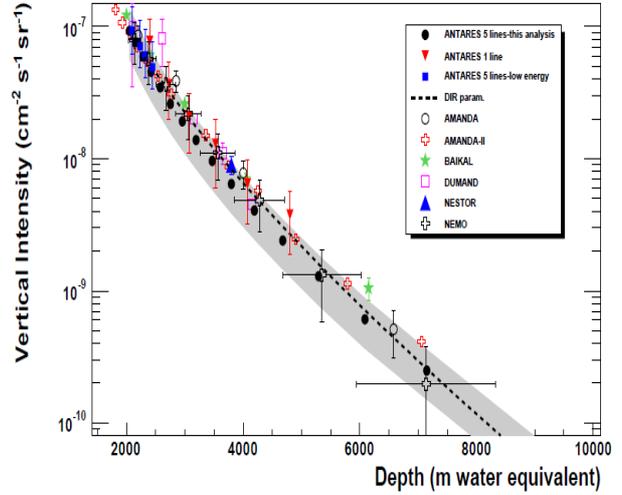

Fig. 4. Compilation of vertical muon intensity data of atmospheric muons vs equivalent water depth.

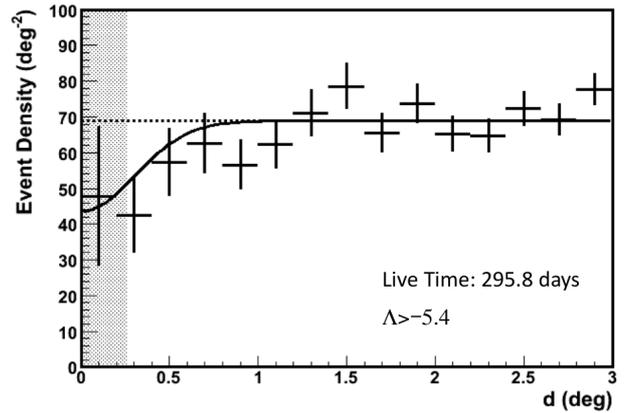

Fig. 5. Preliminary ANTARES muon data on the moon shadow.

Two different searches were performed on the data. The first search was a full sky survey with no assumptions about the source position. The second search was made for a signal excess in a priori fixed spots in the sky corresponding to the positions of some interesting astrophysical objects.

A sky map in galactic coordinates of these events is shown in Fig. 6 [24].

These measurements continue to obtain more information on atmospheric neutrinos [31,10.VI]. The angular resolution is presently 0.5° and may be slightly improved.

A search amongst a predefined list of most promising cosmic sources found in γ ray astronomy can be made. Unfortunately with present statistics we do not find any confirmation with neutrinos above our present background. We can therefore obtain only 90% CL limits as shown in Fig. 7.

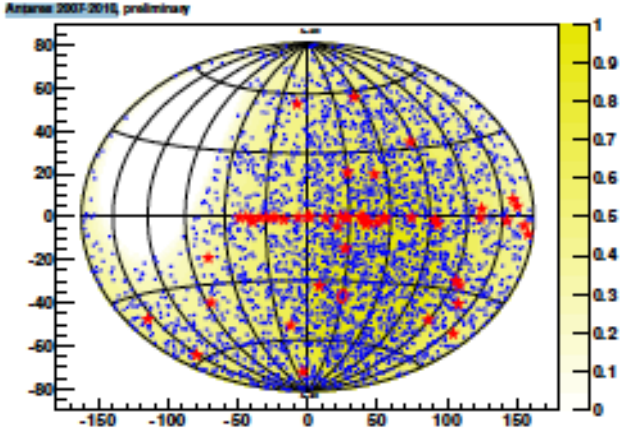

Fig. 6. Galactic skymap showing the 3058 data events. The position of the most signal-like cluster is indicated by a circle. The red stars denote the positions of the 51 candidate γ sources   ( 2007-2010 neutrino data).

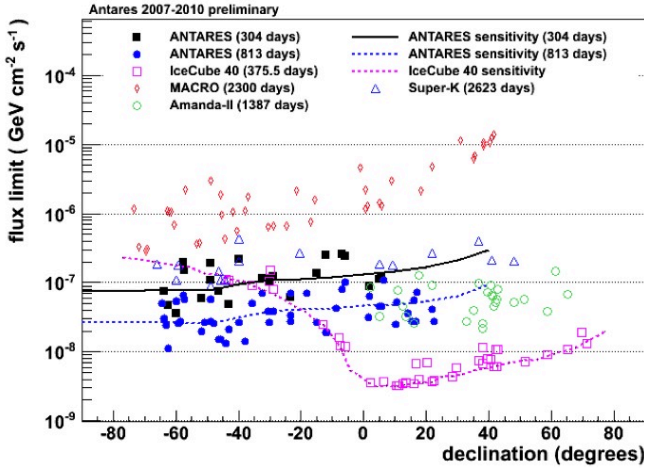

Fig. 7. 90% CL neutrino flux upper limits on an $E^{-2}$ spectrum of high energy neutrinos from 51 selected candidate cosmic sources. Also shown is the sensitivity, which is defined as the median expected limit. In addition to the present result, several previously published limits on sources in both the Southern and Northern sky are obtained by ANTARES (red dots) and other experiments ( MACRO [29], (AMANDA [32] and SuperK [33] ).

## 5. HIGH ENERGY DIFFUSE NEUTRINO FLUX

The data were collected from December 2007 till December 2009 for a total livetime of 334 days. The runs had periods with 9, 10 or 12 lines in operation. Selected runs had a baseline <120 kHz and a burst fraction <40% [34]. The reconstruction algorithm yielded approximately 5% downgoing muons reconstructed as upgoing muons. They were discarded by applying geometrical considerations and with a cut in the reconstruction quality parameter $\Lambda>-5.5$.

A high energy neutrino estimator was used; it is based on the hit repetitions in the OMs due to the different arrival time of *direct* and *delayed* photons. The mean number of repetitions in the event is defined as the number of hits in the same OM within 500 ns from the earliest hit selected by the reconstruction algorithm : $R=R_i/N_{OM}$, where $R_i$ is the number of repetitions in OM i (in most cases $R_i$ is equal to 1, 2), $N_{OM}$ is the number of OMs in which hits used by the reconstruction algorithm are present. Fig. 8 shows the mean number of repetitions R versus the true MC neutrino-induced muon energy. Note that R is approximately linear with log $E_\mu$ in the logaritmic range 3.8-6.

The number of observed events after the proper cuts is compatible with the expected background. One can thus only compute the 90% CL upper limit which at 90% CL for the quantity $E^2 \Phi$ is
$E^2 \Phi_{90\%} = 5.3 \; 10^{-8} \; cm^{-2} \; s^{-1} \; sr^{-1}$
This limit is shown in Fig 9; it is valid in the energy range between 20 TeV and 2.5 PeV ; the limit is compared with other measured flux upper limits [34] and with two model predictions.

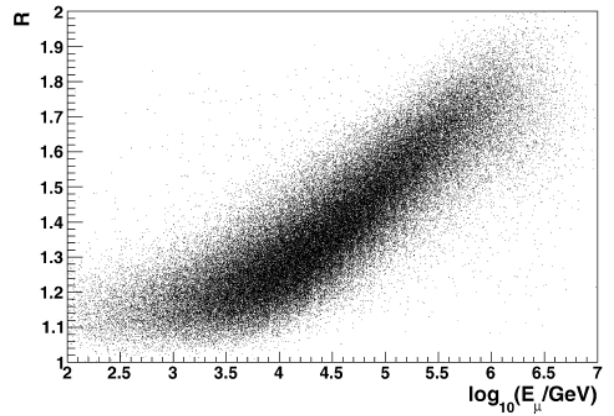

Fig. 8. Mean number of repetitions R as a function of the MC true neutrino-induced muon energy.

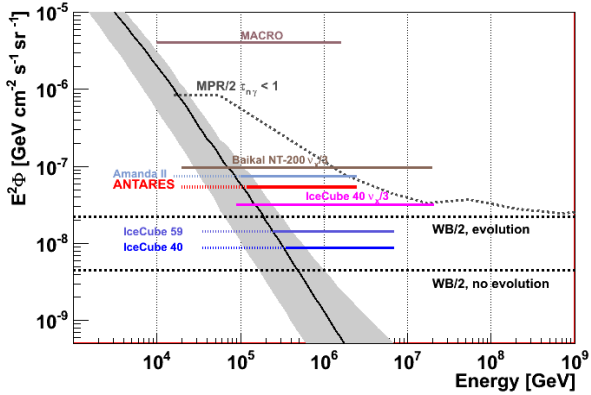

Fig. 9. The ANTARES 90% CL upper limit for an $E^{-2}$ diffuse high energy $\nu_\mu$+antiv$_\mu$ flux. The result is compared with the limits from other experiments. The grey band is the expected variation of the atmospheric $\nu_\mu$ flux and the central red line is averaged over all directions.

## 6. MULTI MESSENGER ASTRONOMY

In order to increase the discovery potentials of ANTARES several collaboration programs with other types of experiments were made. In this "multi messenger" approach the detection thresholds can be lowered in each separate experiment while preserving an acceptable rate of accidental coincidences.

*6.1. Neutrinos from flaring blazars.*
A time dependent point source search was made looking for neutrinos in correlation with variable gamma-ray emissions from blazars observed with the LAT instrument on the Fermi satellite. By restricting the observation to the "high state" gamma emission (lasting typically 1-20 days) the background is reduced compared to the normal search. In an exposure of about 60 days no significant excess was observed [10.VIII].

*6.2. Neutrinos from GRBs.*
Many models predict high energy neutrinos emitted by sources of gamma ray bursts (GRBs). A first search was performed demanding time correlation and angular correlation with GRBs. The search was made using 37 gamma sources when ANTARES had only 5 lines [10.VII]. No coincidence was found. A second method looks for gamma ray shower events produced by electron neutrinos or by neutral current interactions. This search is in progress.

*6.3. Optical follow-up of ANTARES events.*
To search for transient sources of neutrinos with an optical telescope a collaboration was organized to enable fast optical observations in the direction of the detected neutrinos. A fast on line reconstruction algorithm finds the muon track directions and alerts small automatic optical telescopes in Chile and in other south American places. Since 2009 ANTARES sent 37 alert triggers to the TAROT and ROTSE telescope networks in a time window of 15 minutes; 27 signals were followed [10.IX][35].

*6.4. Neutrinos and gravitational wave detectors.*
Another example of such collaboration program is being made with the gravitational wave detectors VIRGO and LIGO. Both detectors had a data-taking phase during 2007, which partially overlapped with the ANTARES 5 line operation. A new common run operation started with the ANTARES 12-line configuration [10.XII].

*6.5. Neutrinos and UHECRs.*
A correlation search was made between Antares neutrinos (from the 2007-2008 run) and the Pierre Auger Observatory 69 UHECR events [10.XI]. The search was negative.

## 7. DARK MATTER SEARCHES

Many theoretical models predict the existence of Weakly Interacting Massive Particles (WIMPs), probably relics from the Big Bang, which may explain the Structures in the Universe and the Dark Matter. In Supersymmetric models (SUSY) the lightest particle may be stable and may correspond to a WIMP.

WIMPs may gravitationally concentrate in the center of massive bodies, like in the center of the Sun, of the Earth and in the center of our Galaxy. If their number is high they may annihilate into normal particles or neutralinos, yielding by decay neutrinos.

Neutrino telescopes may investigate this possibility. In particular ANTARES made a first search when was under construction in 2007-2008 with muon neutrinos coming from the Sun: the search yielded preliminary limits quite dependent on the model chosen [10.XIII][37]. With the full detector and with the statistics already accumulated one is now expecting a better sensitivity. Moreover one has accessible the center of our Galaxy.

## 8. ATMOSPHERIC NEUTRINOS

Measurements of the atmospheric neutrino spectrum are an important test to check the conventional atmospheric neutrino fluxes. Recent studies have shown that a clean sample of atmospheric muon neutrinos with energies as low as 20 GeV can be isolated in the ANTARES neutrino telescope. Such a threshold is low enough to allow the observation of muon neutrino oscillations. A robust analysis method will allow the extraction of atmospheric muon neutrino oscillation parameters [10.VI]. Although not competitive with dedicated experiments [36], it allows to reach a sensitivity to these parameters, which is a demonstration of the understanding of the ANTARES detector. Preliminary analysis of a restricted data sample is compatible with existing constraints on atmospheric neutrino oscillation parameters [37].

## 9. EXOTICA

Neutrino telescopes may also allow dark matter searches [38], searches for exotic particles (fast intermediate mass magnetic monopoles [39], slow nuclearites [10.XVI], etc.), searches for possible violations of general conservation laws [40], atmospheric neutrino studies [3][35], etc.

### 9.1 Magnetic Monopoles (MMs).

MMs are required in many models of spontaneous symmetry breaking. The fast MMs which can be searched for in Neutrino Telescopes like ANTARES cannot be the GUT monopoles which are too massive, but could be the Intermediate Mass Magnetic monopoles with masses $10^{10}$-$10^{14}$ GeV [43]. Fig. 10 shows the limits obtained by ANTARES, which are compared with previous limits obtained by several experiments. In this limited beta range our limits compare well with those previously published [41,42]. And we can do considerably better with more data taking.

### 9.2. Nuclearites.

If Strange Quark Matter (SQM) is the ground state of Quantum Chromodynamics (QCD), nuggets of SQM (Nuclearites) may be present in the penetrating cosmic rays. They could have been produced in the Early Universe or in violent astrophysical processes. A phenomenological bound for the nuclearite flux in our galaxy was derived from the dark matter density.

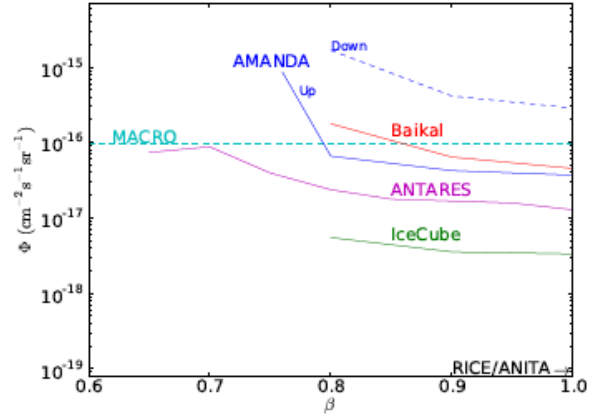

Fig. 10. 90% CL limits on fast intermediate mass magnetic monopoles obtained by ANTARES. They are compared with limits from AMANDA, Baikal, MACRO, ICECUBE and with the Parker theoretical limit.

The nuclearite detection in a large volume neutrino telescope is possible through the black body emission from the overheated path in water [10.XVI]. The chemical potential difference between $s$ and $u$ or $d$ quarks may induce a small residual positive electrical charge for nuclearites, which should travel with the typical velocity of gravitationally trapped objects in our Galaxy, $\beta$=v/c=$10^{-3}$. The residual electric charge should be compensated by an electron cloud and/or by electrons in weak equilibrium inside the SQM. Since one expects that the nuclearite flux in the cosmic rays decreases with increasing mass (as for heavy nuclei) we concentrated on downgoing nuclearites with mass m>$10^{22}$ GeV. In water it is estimated that the nuclearite energy loss is dissipated mainly in a large number of visible photons. The search for nuclearites was based on the fact that the duration of a nuclearite event should be much longer (about 1 ms) than in the case of a muon. A MC simulation was developed and used in the analysis of the data collected in 2007-2008 with a variable number of lines. No event passed all the cuts and the visual scanning and we can only quote in Fig. 11 the 90% CL limits obtained [10.XVI].

## 10. ACOUSTIC DETECTION

Because of the long attenuation length of acoustic signals, about 5 km for 10 kHz signals, the detection of acoustic pressure pulses in large underwater detectors may be a good method to detect cosmic neutrinos with energies larger than 100 PeV. The ANTARES infrastructure incorporates the AMADEUS acoustic system for the calibration of the detector and for detecting the acoustic background in the sea [18]. The sensors are distributed in "acoustic clusters" installed along line 1 and the IL07 line at a horizontal distance of 240 m; the vertical spacing of the sensors in a line range from 15 m to 125 m. Because of the surface waves generated by winds the noise level is larger than in the lab measurements. An acoustic sea map shows the detected signals coming from the calibration signals, from the noise generated by surface boats, from surface waves and from marine mammals.

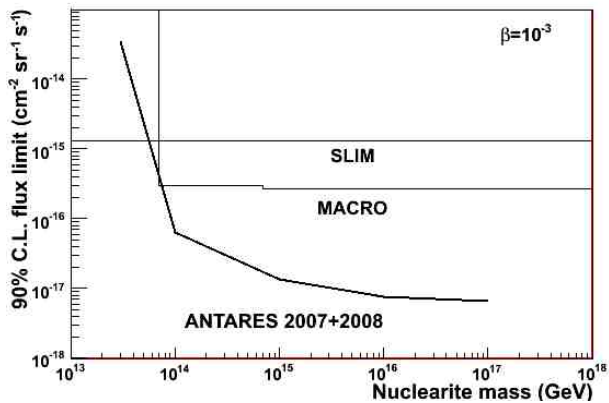

Fig. 11 . The ANTARES 90% CL upper limits for a flux of downgoing nuclearites obtained from the 2007 and 2008 data. The limits are compared with those obtained by the MACRO [41] and SLIM [42] experiments.

11. EARTH AND SEA SCIENCE

Earth and Sea Science is a large interdisciplinary field. ANTARES is well equipped for research in this field and the first preliminary results were obtained using an Acoustic Doppler Current Profiler (ADCP) and the optical PMTs [11].

*11.1. Oceanographic and Bioluminescence studies*. ANTARES offers the facilities to compare high resolution acoustic and optical observations, 70 m and 170 m above the sea bed. The ADCP measured downward vertical currents particularly important in late winter and early spring. In the same period observations were made of enhanced levels (by a factor of about 19) of acoustic reflections interpreted as due to suspended particles, including zooplankton, and horizontal water currents reaching 35 cm/s . The observations coincided with high light levels detected by the PMTs of the telescope, interpreted as increased bioluminescence. During winter 2006 deep dense water formations occurred, providing a possible explanation for these observations. It was hypothesized that the main process for suspended material to be moved vertically later in the year is linked with topographic boundary current instabilities along the rim of the Northern Current in the Mediterranean sea, Fig.12.

*11.2. Underwater seismometer*. Fig. 13 shows the readings from our underwater seismometer during the last great earthquake off the east coast of Japan (plus the terrible tsunami). The Earthquake is well visible. Clearly the main purposes of an underwater seismometer are somewhat different, but one may say that one obtained a calibration of the instrument.

11.3. ANTARES is now equipped also with an *underwater camera* which can show the luminous marine life, luminous bacteria and larger size luminous marine life around the telescope.

11.4. In some of the ANTARES lines there is a very good instrumentation to measure with high precision the salinity and temperature of the bottom layer of the sea. And there are good instruments to measure the speed of the deep water currents, both in the horizontal plane and also in the vertical direction. Analyses indicated correlations of the light noise from luminous marine bacteria and from larger marine life with the sea currents, in particular also with vertical displacements of marine layers.

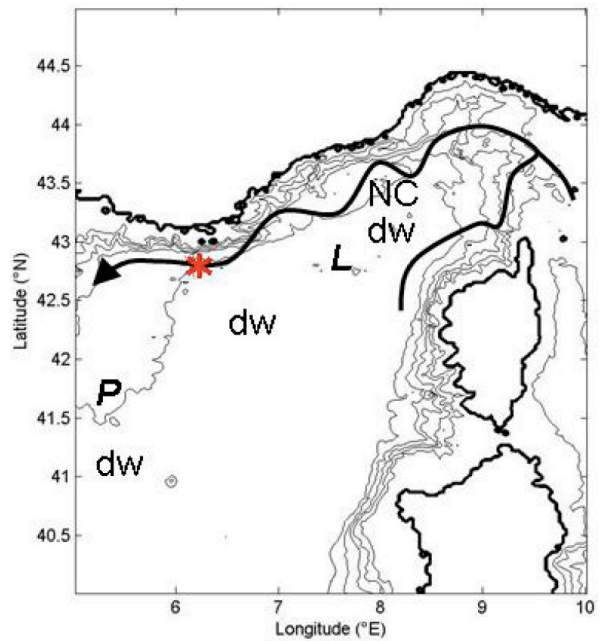

Fig. 12. ANTARES site (red star) located on the northern Edge of the border between the Ligurian (L) and Provencal (P) sub-basins. Also indicated are the Northern current (NC) and areas of dense water formations (dw).

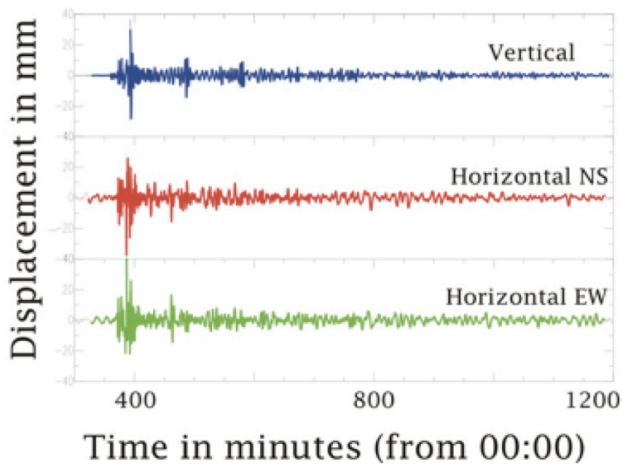

Fig. 13. The earthquake of March 11, 2011, of very large magnitude in Japan, observed at the ANTARES site.

## 12. CONCLUSIONS. OUTLOOK

The ANTARES neutrino telescope is running in its final 12 line configuration and is taking also data with a variety of instruments for Earth and Sea science research.

The data on downgoing muons were used to calibrate the detector and to measure the vertical muon intensity as a function of the water depth. There is good agreement with previous data and with MC expectations. The 30% systematic uncertainty is mainly connected with an uncertainty in the primary cosmic ray model [26].

The first results of the search for cosmic point sources of neutrinos were obtained with an angular resolution of about 0.5°. ANTARES is now analyzing more data and is trying to search for intermittent sources with multi messenger methods, in collaboration with other types of detectors [10].

A search was also made for a diffuse flux of high energy (HE) neutrinos.

Of particular interest is the study of HE neutrinos from the center of our galaxy and from the Sun.

Good limits were obtained in the search for fast Magnetic Monopoles and slow Nuclearites.

All the above searches will be repeated with higher statistics.

The successful operation of ANTARES and the many analyses results obtained are important steps forward to Km3Net, a future deep sea $km^3$ neutrino observatory and marine sciences infrastructure planned for construction in the Mediterranean Sea.

## Acknowledgements

I acknowledge the advice, cooperation and technical support from many colleagues, in particular S. Biagi, S. Cecchini, T. Chiarusi, P. Coyle, A. Margiotta , G. Pavalas, M. Spurio and J.D. Zornoza.

## References

1. J. Babson et al., Cosmic ray muons in the deep ocean, Phys. Rev. D 42 (1990) 3613-3620.
2. I. A. Belolaptikov et al., The Baikal underwater neutrino telescope, Astropart. Phys.7( 1997) 263-282.
3. K. Nakamura et al., Review of Particle Physics, J. Phys. G 37 (2010) 1.
4. R. Abbasi et al., Search for a Lorentz-violating sidereal signal with atmospheric neutrinos in Ice Cube, Phys. Rev. D82 (2010) 112003.
5. G. Aggouras et al., A measurement of the cosmic-ray muon flux with a module of the NESTOR neutrino telescope, Astropart. Phys. 23 (2005) 377-392.
6. E. Migneco et al., Recent achievements of the NEMO project, Nucl. Instrum. Meth. A588 (2008) 111-118.
7. M. Ageron et al., ANTARES: the first undersea neutrino telescope, Nucl. Instrum. Meth. A656 (2011) 11-38.
8. Km3Net : http://www.km3net.org/home.php
9. G. Giacomelli, Results from the ANTARES neutrino telescope, arxiv:1105.1245 [astro-ph-IM].
10. S. Adrian-Martinez et al. (The ANTARES Collaboration: contributions to ICRC2011)Beijing,arXiv:1112.0478 [astro-ph.HE], In the text the ref. is indicated as [10.Section in roman number].
11. H. van Haren et al, Acoustic and optical variations during rapid downward motion episodes in the deep north-western Mediterranean sea, Deep-Sea Research I 58 (2011) 875-884.
12. J. A. Aguilar et al., Transmission of light in deep sea water at the site of the Antares neutrino telescope, Astropart. Phys. 23 (2005) 131-155.
13. S. Adrian Martinez et al., Measurement of the group velocity of light in sea water at the ANTARES site, Astropart. Phys. 35 (2012) 552-557.
14. P. Amram et al., Sedimentation and fouling of optical surfaces at the Antares site, Astropart. Phys. 19 (2003) 253-267.
15. P. Amram et al., The ANTARES optical module, Nucl. Instrum. Meth. A 484 (2002) 369-383.
16. J. A. Aguilar et al., Study of large hemispherical photomultiplier tubes for the ANTARES neutrino telescope, Nucl. Instrum. Meth. A555 (2005) 132-141.
17. J. A. Aguilar et al., Time calibration of the ANTARES neutrino telescope, Astropart. Phys 34 (2011) 539-549.
18. J. A. Aguilar et al., AMADEUS-The acoustic neutrino detection test system of the ANTARES deep-sea neutrino telescope, Nucl. Instrum. Meth. A626-627 (2011) 128-143.

19. M. Ageron et al., The Antares optical beacon system, Nucl. Instrum. Meth. A578 (2007) 498-509.
20. J. A. Aguilar et al., First results of the instrumentation line for the deep-sea ANTARES neutrino telescope, Astropart. Phys. 26 (2006) 314-324.
21. M. Ageron et al., Studies of a full scale mechanichal prototype line for the ANTARES neutrino telescope, Nucl. Instrum. Meth. A581 (2007) 695-708.
22. J. A. Aguilar et al., The data acquisition system for the ANTARES neutrino telescope, Nucl. Instrum. Meth. A570 (2007) 107-116.
23. J. A. Aguilar et al., Performance of the front-end electronics of the ANTARES neutrino telescope, Nucl. Instrum. Meth. A622 (2010) 59-73.
24. M. Ageron et al., Performance of the first ANTARES detector line, Astropart. Phys. 31 (2009) 277-283.
25. J. A. Aguilar et al., Measurement of the atmospheric muon flux with a 4 GeV threshold in the ANTARES neutrino telescope, Astropart. Phys. 33 (2010) 86-90.
26. J. A. Aguilar et al., Zenith distribution and flux of atmospheric muons measured with the 5 line ANTARES detector, Astropart. Phys. 34 (2010) 179-184.
27. M. Ambrosio et al., Vertical muon intensity measured with MACRO at the Gran Sasso laboratory, Phys. Rev. D52 (1995) 3793-3802.
28. M. Ambrosio et al., Moon and sun shadowing effect in the MACRO detector, Astropart. Phys. 20 (2003) 145-156.
29. M. Ambrosio et al., Neutrino astronomy with the MACRO detector, Astrophys. J. 546 (2001) 1038-1054.
30 S. Adrian-Martinez et al., First search for point sources of high energy cosmic neutrinos with the ANTARES detector, Astrophys. J. Lett. 743 (2011) L14-L19.
31. J. A. Aguilar et al., A fast algorithm for muon track reconstruction and its application to the ANTARES neutrino telescope, Astropart. Phys 34 (2011) 652-662.
32. A. Kappes, News from the south pole: Recent results from IceCube and AMANDA, Nucl. Phys. A827 (2009) 567C-569C.
R. Abbasi et al., IceCube Astrophysics and Astroparticle physics at the south pole, arXiv:1111.5188 [astro.ph-HE].
33. E. Thrane et al., Search for astrophysical neutrino point sources at Super-K, Astrophys. J. 704 (2009) 503-512.
34. J. A. Aguilar et al., Search for a diffuse flux of high energy $\nu_\mu$ with ANTARES, Phys. Lett. B696 (2011) 16-22.
G. Sullivan, for the IceCube Collab., Neutrino 2012 Conf., Results from the IceCube experiment, arXiv:1210.4195 [astro.ph-HE].
35. M. Ageron et al., The ANTARES telescope neutrino alert system, Astropart. Phys. 35 (2012) 530-536.
36. M. Ambrosio et al., Measurement of the atmospheric neutrino-induced upgoing muon flux using MACRO, Phys. Lett. B434 (1998) 451-457; Matter effects in upper going muons and sterile neutrino oscillations, Phys. Lett. B517 (2001) 59-66.
Y. Fukuda, et al., Evidence for oscillation of atmospheric neutrinos, Phys. Rev. Lett. 81 (1998) 1562-1567.
Q.R Ahmad et al., Measurement of the rate of nu/e + d --> p + p + e- interactions produced by B-8 solar neutrinos at the Sudbury Neutrino Observatory, Phys. Rev. Lett. 87 (2001) 071301.
P. Adamson et al., Measurement of neutrino oscillations with the MINOS detectors in the NuMI beam, Phys. Rev. Lett. 101 (2008) 131802.
37. S. Adrian-Martinez et al, Measurement of neutrino oscillation with the ANTARES neutrino telescope, Phys. Lett. B714 (2012) 224, arXiv:1206.0645 [hep-ex].
38. G. Lim, First results on the search for dark matter with ANTARES, arXiv:0905.2316 v3 [astro-ph.CO].
IceCube Collab., Search for DM annihilation in the Sun with the 79-string detector, arXiv:1206.0645 [hep-ex].
39. S. Adrian Martinez et al., Search for relativistic magnetic monopoles with ANTARES, arXiv:1110.2656 [astro-ph.HE], Astropart. Phys 35 (2012) 634-640.
R. Abbasi et al., Search for relativistic magnetic monopoles with IceCube, arXiv:1208.4861 [astro.ph-HE].
40. G. Battistoni et al., Search for a Lorentz invariance violation in atmospheric neutrino oscillationa using MACRO data, Phys. Lett. B615 (2005) 14-18.
41. M. Ambrosio et al., Final results of magnetic monopole searches with the MACRO experiment, Eur. Phys. J. C25 (2002) 511-522.
42. S. Cecchini et al., Search for intermediate mass magnetic monopoles and nuclearites with the SLIM experiment, Radiat. Meas. 40 (2005) 405-409.
43. T.W.Kephart and Q. Shafi, Family unification, exotic states and magnetic monopoles, Phys. Lett. B520 (2001) 313-316.